\documentclass[a4paper,USenglish,cleveref, autoref, thm-restate]{lipics-v2021}
\pdfoutput=1
\graphicspath{{./graphics/}}

\title{Spanner Approximations in Practice}
\titlerunning{Spanner Approximations in Practice}

\author{Markus Chimani}{Theoretical Computer Science, Universität Osnabrück, Germany}{chimani@uos.de}{https://orcid.org/0000-0002-4681-5550}{}
\author{Finn Stutzenstein}{Theoretical Computer Science, Universität Osnabrück, Germany}{fistutzenste@uos.de}{}{}

\authorrunning{M. Chimani, F. Stutzenstein}

\Copyright{Markus Chimani, Finn Stutzenstein}



\category{} %optional, e.g. invited paper

\relatedversion{} %optional, e.g. full version hosted on arXiv, HAL, or other respository/website
\nolinenumbers %uncomment to disable line numbering

\hideLIPIcs  %uncomment to remove references to LIPIcs series (logo, DOI, ...), e.g. when preparing a pre-final version to be uploaded to arXiv or another public repository

\EventEditors{John Q. Open and Joan R. Access}
\EventNoEds{2}
\EventLongTitle{42nd Conference on Very Important Topics (CVIT 2016)}
\EventShortTitle{CVIT 2016}
\EventAcronym{CVIT}
\EventYear{2016}
\EventDate{December 24--27, 2016}
\EventLocation{Little Whinging, United Kingdom}
\EventLogo{}
\SeriesVolume{42}
\ArticleNo{23}
\usepackage{makecell}
\usepackage{xspace}
\usepackage[table]{xcolor}
\renewcommand{\O}{\mathcal{O}}
\newcommand{\ADDJS}{\texttt{ADDJS}\xspace}
\newcommand{\KP}{\texttt{KP}\xspace}
\newcommand{\BBMRY}{\texttt{BBMRY}\xspace}
\newcommand{\BS}{\texttt{BS}\xspace}
\newcommand{\EN}{\texttt{EN}\xspace}

\usepackage[shortcuts]{extdash}
\usepackage[ruled,vlined]{algorithm2e}
\usepackage{diagbox}
\usepackage{relsize}
\usepackage{multirow}

\newcommand{\mypar}[1]{\smallskip\noindent\textsf{\textbf{#1~}}}

\begin{document}

\maketitle

\begin{abstract}
	A multiplicative $\alpha$-spanner $H$ is a subgraph of $G=(V,E)$ with the same vertices and fewer edges that preserves distances up to the factor $\alpha$, i.e., $d_H(u,v)\leq\alpha\cdot d_G(u,v)$ for all vertices $u$, $v$. While many algorithms have been developed to find good spanners in terms of approximation guarantees, no experimental studies comparing different approaches exist. We implemented a rich selection of those algorithms and evaluate them on a variety of instances regarding, e.g., their running time, sparseness, lightness, and effective stretch.
\end{abstract}

\section{Introduction}\label{sec:introduction}
Given a directed or undirected graph $G=(V,E)$ with $n$ vertices and $m$ edges, the shortest distance between two vertices $u$ and $v$ is denoted by $d_G(u,v)$. A graph spanner is a sparse subgraph of a given graph that preserves these distances to some quality degree. Spanners were introduced by Peleg and Schäffer \cite{Peleg1989} after the first mention by Awerbuch \cite{Awerbuch1985}. There are many spanner variants, see Ahmed et. al \cite{Ahmed2020} for a survey. In this paper we are going to focus on the probably most popular variant:
\begin{definition}[Multiplicative $\alpha$-Spanner \cite{Peleg1989}]\label{def:spanner}
	Given a directed or undirected graph $G=(V,E)$ and a \emph{stretch} $\alpha\geq 1$, a multiplicative $\alpha$-spanner $H=(V,E')$ is a subgraph of $G$ with $E'\subseteq E$ such that $d_H(u,v)\leq\alpha\cdot d_G(u,v)$ holds for all pairs $(u,v)\in V\times V$.
\end{definition}
For readability we are going to use the term \emph{spanner} for the multiplicative $\alpha$-spanner in the following. The definition is also applicable if $G$ has additional edge weights $w(e)>0$.

Given a graph $G$, finding any spanner is trivial since $G$ is a spanner of itself. The problem we are interested in is to find a \emph{good} spanner. There are three basic measures of a spanner's quality: First, the \emph{size} referring to the number of edges $|E'|$. The \emph{sparseness} $s(H)=\frac{|E'|}{|E|}$ is the relative size w.r.t.\ the original graph, with a lower value indicating a sparser spanner. Lastly, the \emph{lightness} $\ell(H)=\frac{W(E')}{W(\mathrm{MST}(G))}$ with $W(F)=\sum_{e\in F}w(e)$ compares the weight of the spanner to the weight of the original graph's minimal spanning tree. This results in the problem of finding the \emph{sparsest} or \emph{lightest} spanner:
\begin{definition}[Sparsest (Lightest) $\alpha$-Spanner]\label{def:spanner-problem}
	Given a graph $G=(V,E)$ and a stretch $\alpha\geq1$, find a sparsest (lightest) multiplicative $\alpha$-spanner $H=(V,E')$ such that $s(H)\leq s(H')$ ($\ell(H)\leq \ell(H')$, respectively) for all other spanners $H'$ of $G$.
\end{definition}

The problems of finding the sparsest and lightest spanner are NP-hard \cite{Peleg1989, Cai1994}. A variety of algorithms were developed to tackle the problem of finding good spanners, motivated by the diverse needs of applications: Several routing problems use spanners \cite{Awerbuch1985, Peleg1989application, Cowen2001, Shpungin2009}. In distributed computing setups, spanners can improve the broadcast of messages between nodes \cite{Peleg1987, Cowen2004}. Also many theoretical applications themselfs require spanners, like approximate distance oracles \cite{Peleg1999, Thorup2005}, almost shortest paths \cite{Elkin2005, Elkin2006}, or access control hierarchies \cite{Jha2003, Bhattacharyya2009}.

We are interested in the computation of spanners in practice. In 2004, Sigurd and Zachariasen \cite{Sigurd2004} developed an exact algorithm based on an ILP formulation with an exponential number of variables, each representing a shortest path. Graphs with up to 64 vertices were tested and not all instances could be solved within the timelimit of 30 minutes. Ahmed et al.\ \cite{Ahmed2019} recently published a compact ILP formulation based on a multicommodity flow, i.e., it has only a polynomially number of variables and constraints. The ILP is solved by standard Branch-and-Bound. They tested graphs with up to 100 vertices; the solver needed up to 40 hours to solve the largest instances on a high-performance 400-node cluster. They did not compare the results to \cite{Sigurd2004} or other popular heuristics. Apart from the exact algorithms, Farshi und Gudmundsson \cite{Farshi2005} did some experimental studies for the special case of geometric spanners. To our best knowledge, no further published experimental study exists.

\mypar{Contribution.}
Since almost no experimental studies exist on this topic despite the multitude of applications, we aim to close this gap and provide first experimental results for a variety of popular non-exact algorithms in the general, non-geometric setting. This leads to the further layout of this paper: We provide an overview over the published algorithms in \cref{sec:previous-work}. We argue our selection of algorithms and give implementation details in \cref{sec:implementation-details} and we present the results of our experimental study in \cref{sec:experimental-results}.

\section{Previous Work}\label{sec:previous-work}
Much work and effort has been put into developing different approaches to find spanners. We will categorize the existing literature into three main categories: greedy algorithms, approximation algorithms, and other methods. For a rich theoretical survey of graph spanners, their theoretical background and algorithms, see Ahmed et al.\ \cite{Ahmed2020}.

\mypar{Greedy algorithms.}
One of the earliest and simplest algorithms is by Althöfer et al.\ \cite{Althoefer1993} and is reminiscent of Kruskal's algorithm. The algorithm \ADDJS, often named \emph{Basic Greedy Spanner Algorithm}, starts with a spanner without edges and adds edges $\{u,v\}$ with increasing weight to the spanner, if the shortest path between $u$ and $v$ is currently too long, i.e., $d_H(u,v)>\alpha\cdot w(\{u,v\})$. For a stretch $\alpha=2k-1$ ($k\in\mathbb{N}_{\geq1}$), this algorithm creates a spanner of size $\O(n^{1+1/k})$ and lightness $\O(n/k)$. The running time is mostly dominated by a Dijkstra run for each edge: $\O(m(n^{1+1/k}+n\log n))$.

%\begin{algorithm}[t]
%	\SetKwFunction{Sort}{Sort}
%	\SetKwInOut{Input}{Input}\SetKwInOut{Output}{Output}
%	\Input{$G=(V, E)$, weights $w(e)$, $\alpha\geq1\in\mathbb{R}$}
%	\Output{An $\alpha$-spanner $H=(V,E')$}
%	
%	$E'\leftarrow\emptyset$\;
%	Sort $E$ by $w(e)$ in ascending order\;
%	\For {all $e=\{u,v\}\in E$} {
%		\If {$d_H(u,v)>\alpha\cdot w(e)$} {
%			$E'\leftarrow E'\cup\{e\}$\;
%		}
%	}
%	
%	\caption{\ADDJS algorithm.}
%	\label{list:addjs}
%\end{algorithm}

Many publications aim to improve \ADDJS to provide better guarantees for either the size, lightness, or running time. They all have a stretch of $\alpha=2k-1$ in common (except for some $\varepsilon$-parameterized algorithms) and take undirected, weighted graphs as inputs: Roditty and Zwick \cite{Roditty2004} improved the distance calculation in the spanner by incrementally maintaining a shortest-path tree. This speeds up the algorithm, but does no longer provide a lightness guarantee for the resulting spanner. Thorup and Zwick \cite{Thorup2005} explored the topic of \emph{approximate distance oracles} which resulted in a spanner algorithm as a byproduct. Similar to before, no lightness guarantee can be given. Chandra et al.\ \cite{Chandra1992} and Elkin and Solomon \cite{Elkin2016} use various techniques (e.g., dynamic shortest-path trees, further auxillary graphs, and distance oracles) to create a spanner with guarantees for size and lightness as well as maintaining a low running time. Alstrup et al.\ \cite{Alstrup2019} further improved these results to provide the currently best size guarantee. The algorithm creates a $((2k-1)(1+\varepsilon))$-spanner with size $\O_\varepsilon(n^{1+1/k})$ and lightness $\O_\varepsilon(n^{1/k})$ within a running time of $\O(n^{2+1/k+\varepsilon'})$. Here, $\O_\varepsilon$ ignores polynomial factors depending on $1/\varepsilon$. However, \ADDJS still has the best lightness guarantee. Additionally, most modifications lead to very complex, often not practically implementable algorithms. We will thus only consider the original \ADDJS in our experimental study.

\mypar{Approximation algorithms.}
An early $\O(\log(m/n))$-approximation in terms of size by Kortsarz and Peleg \cite{Kortsarz1994} was published in 1994 for the special case of a 2-spanner for unweighted and undirected graphs (algorithm \KP). An edge $e$ is \emph{covered}, if it is not part of the spanner, but part of a triangle in the original graph and both other edges are in the spanner, so the spanner property holds for $e$. The idea is to cover a large number of edges while not adding too many edges to the spanner by finding dense subsets of vertices and adding connecting edges to the spanner. \KP calculates \emph{dense subsets} $U_v$ of neighboring vertices $N(v)$ for each vertex $v$ and continues with the densest subset, say $U_w$. The star $\{\{u,w\}|u\in U_w\}$ is added to the spanner. The edges within the dense subset are now covered and finding the densest subgraph allows for covering the most edges. This is repeated while there exists a dense subset with a density larger than 1. The running time is bounded by $\O(nm\cdot\mathrm{MDS}(n, m))$ where $\mathrm{MDS}(n, m)$ is the running time of the algorithm to solve the maximum density subgraph problem. Using Goldberg's algorithm \cite{Goldberg1984}, the resulting time complexity is $\O(m^2n^3\log n)$. There exist more advanced algorithms for the maximum density problems so the theoretical running time can be lowered to $\O(m^2n^2\log (n^2/m))$ \cite{Gallo1989}.

For undirected graphs we know of no explicit further approximation results. One reason may be that \ADDJS already has a $\O(n^{1+1/k})$ guarantee for the spanner size. For a connected original graph, a spanner must have at least $n-1$ edges, as it has to be connected as well. This lower bound for an optimal solution thus results in a straight-forward $\O(n^{1/k})$\=/approximation in terms of size for every undirected graph.

For directed graphs, Elkin and Peleg \cite{Elkin2005} gave the first $\tilde{\O}(n^{2/3})$-approximation in the special case $\alpha=3$. For general $\alpha\geq3$, Bhattacharyya et al.\ \cite{Bhattacharyya2009} published an $\tilde{\O}(n^{1-1/\alpha})$-approximation combining two techniques: The spanner is the union of the arc sets of two graphs. The first graph is obtained by solving an LP relaxation and rounding the solution. The second one is created by sampling vertices and growing BFS arborescences from them. Dinitz and Krauthgamer \cite{Dinitz2010} improved the techniques and Berman et al.\ \cite{Berman2013} gave the currently best solution for general stretch $\alpha\geq1$ (algorithm \BBMRY): They introduced an $\O(n^{1/2}\log n)$-approximation for weighted, directed graphs. For the special case $\alpha=3$ (later extended to $\alpha=4$ by Dinitz and Zhang \cite{Dinitz2016}) a slightly modified approach can be taken to achieve an $\tilde{\O}(n^{1/3})$-approximation for directed and unweighted graphs. The techniques used are similar to the approach of Bhattacharyya et al.\ \cite{Bhattacharyya2009}: The arcs of the original graphs are categorized as either \emph{thin} or \emph{thick} by the number of shortest paths connecting the endpoints of the arc. A LP rounding covers all thin arcs and growing arborescences from sampled vertices cover all thick arcs. For the (I)LP, \emph{antispanners} were introduced. Given an arc $(u,v)$, an antispanner is an arc set~$A$, such that no shortest path with a distance less then $\alpha\cdot d_G(u,v)$ exists between $u$ and $v$ in $G\setminus A$. $A$ is a minimal antispanner for $(u,v)$ if no $A'\subset A$ is an antispanner for $(u,v)$. The ILP tries to cover arcs so that at least one arc of each minimal antispanner is used. If this is achieved, the selected arcs resemble a feasible solution. Due to the exponential number of minimal antispanners, we require a separation oracle to solve the LP relaxation of this ILP with cutting planes. The separation oracle has an exponentially small probability in $n$ of failing. Overall, the techniques used are generally suitable for an implementation.
 
\mypar{Alternative methods.}
Some alternative algorithms follow distinctly different ideas. Baswana and Sen \cite{Baswana2007} gave a method based on growing clusters around sampled vertices. This algorithm \BS is special compared to all other algorithms as it does not do any local or global distance calculation. It computes an $\alpha=2k-1$ spanner for a weighted, undirected graph in two phases. First, $k-1$ rounds are done growing clusters by sampling vertices and successively adding nearest neighboring vertices to each cluster. After this phase, clusters are joined by adding all non-clustered vertices to the nearest adjacent cluster and interconnecting all clusters. This yields a spanner of size $\O(kn^{1+1/k})$ in an expected running time of $\O(km)$, but no lightness guarantee. For an unweighted graph, the size is bounded by $\O(n^{1+1/k}+kn)$.

A probabilistic method was presented by Elkin and Neiman \cite{Elkin2018} to calculate a spanner with stretch $2k-1$ for undirected, unweighted graphs (algorithm \EN). They built upon Miller et al.\ \cite{Miller2015} and improved the size to $\O(n^{1+1/k}/\varepsilon)$. For each vertex $u$, a random value $r_u$ is drawn from an exponential distribution. Each vertex broadcasts a message with $r_u$ as the content to all vertices within distance $k$. A vertex $x$, receiving messages from some vertices $U\subset V$, stores $m_u(x)=r_u-d_G(u,x)$ for each vertex $u\in U$ together with an edge adjacent to $x$ that lies on a shortest path between $u$ and $x$. The edges belonging to the messages with $m_u(x)\geq\max_{u\in V}\{m_u(x)\}-1$ are added to the spanner. Since the algorithm has a success probability of at least $1-\varepsilon$, we can obtain an expected running time of $\O(m)$ by iterating.

\section{Considered Algorithms and Implementation Details}\label{sec:implementation-details}
\begin{table}[t]
	\setlength\extrarowheight{5pt}
	\centering
	\smaller
	\begin{tabular}{|c|c|c|c|c|c|c|}\hline 
		Algorithm & & $w(e)$ & Stretch & Spanner Properties* & Running time \\ \hline
		\ADDJS \cite{Althoefer1993} & G & \checkmark & $\alpha\in\mathbb{R}_{\geq 1}$ & \makecell{$s(H)\in\O(n^{1+1/k})$ \\ $\ell(H)\in\O(n/k)$} & $\O(m(n^{1+1/k}+n\log n))$ \\ \hline
		\KP \cite{Kortsarz1994} & A & $\times$ & $\alpha=2$ & $\O(\log(m/n))$-approx.\,in\,size & $\O(m^2n^3\log n)$ \\ \hline
		\BBMRY \cite{Berman2013} & A & \checkmark & $\alpha\in\mathbb{R}_{\geq 1}$ & $\O(n^{1/2}\log n)$-approx.\,in\,size & poly \\ \hline
		\BS \cite{Baswana2007} & C & $\times$ & $\alpha\in\mathbb{N}_{\geq3}$, odd & $s(H)\in\O(n^{1+1/k}+kn)$ & expected $\O(km)$ \\ \hline
		\BS \cite{Baswana2007} & C & \checkmark & $\alpha\in\mathbb{N}_{\geq3}$, odd & $s(H)\in\O(kn^{1+1/k})$ & expected $\O(km)$ \\ \hline
		\EN \cite{Elkin2018} & P & $\times$ & $\alpha\in\mathbb{N}_{\geq1}$, odd & \makecell{$s(H)\in\O(n^{1+1/k}/\varepsilon)$ \\ $0<\varepsilon<1$} & \makecell{expected $\O(m)$, \\ success prob. $\geq1-\varepsilon$} \\ \hline
	\end{tabular}
	\caption{Considered algorithms. The letters G, A, C and P in the second column stand for greedy, approximation, clustering, and prababilistic, respectively. *Assume $\alpha=2k-1$, $k\in\mathbb{N}_{\geq1}$.}
	\label{table:algorithms}
\end{table}

After the overview of different approaches one may wonder which algorithms to use to calculate a good spanner in practice. To close the \emph{Algorithm Engineering Cycle} \cite{MHS2010, Kliemann2016} with implementations and practical results we select five algorithms to analyze. \cref{table:algorithms} provides an overview of the characteristics of each algorithm.

With the goal of small calculation times and larger instances, we decided against using any exact algorithm as they would not generate a feasible solution for the majority of instances. For the greedy algorithms, we selected \ADDJS. While the algorithms of Thorup and Zwick \cite{Thorup2005} as well as Roditty and Zwick \cite{Roditty2004} improve the running time, \ADDJS promises to  produce better results w.r.t.\ size and lightness. The more advanced algorithms are not well implementable as such. From the approximation algorithms we selected \KP and \BBMRY as they are the state-of-the-art. For \BBMRY, we implemented the variant for arbitrary $\alpha$, not the special improvements for $\alpha=3$ and $\alpha=4$. Lastly, \BS and \EN are implemented due to a comparatively low running time and simplicity of the algorithms themselves.

In the following we list some implementation specific details for each algorithm.

\mypar{Althöfer et al.\ (\ADDJS).}
We included a straightforward observation to speed up the computation: We can bound the shortest-path search for each pair of vertices in the spanner by $\alpha\cdot w(\{u,v\})$ since finding a longer path (or no path at all) results in adding the edge~$\{u,v\}$ to the spanner. Note that one is not limited to a stretch $\alpha=2k-1$; the algorithm works for arbitrary $\alpha\geq1\in\mathbb{R}$, but may only provide the size and lightness guarantees for the next odd integer.

\mypar{Kortsarz and Peleg (\KP).}
The algorithm mainly depends on an implementation for the \emph{maximum density subgraph problem}. We use Goldberg's algorithm \cite{Goldberg1984} due to its simplicity of implementation. It resembles a binary search with $\O(\log n)$ steps: A source and target vertex ($s$, $t$) are added to the graph and connected to all original vertices. In each step the weights of the new edges are chosen depending on the current search value. A minimum $s$-$t$-cut is calculated and based on the resulting cut it is decided, whether the lower or upper half of the search interval is used. To solve the minimum $s$-$t$-cut problem we use Goldberg-Tarjan's max-flow algorithm \cite{Goldberg1988} with global relabeling and gap relabeling heuristic requiring $\O(n^2m)$ time. This results in a running time for Goldberg's maximum density subgraph algorithm of $\O(n^2m\log n)$ and, as described above, in a complete running time of $\O(m^2n^3\log n)$ for \KP.

The algorithm has to maintain three sets $H^s$, $H^c$, and $H^u$ for the \textbf{s}panner edges, \textbf{c}overed edges, and \textbf{u}ncovered edges. In the implementation it is sufficient to only use the original graph's edge set $E=H^u$ (by shrinking $G$ on the fly) and the spanner's edge set $E'=H^s$. There is no need to explicitly store covered edges $H^c$ in some data structure. For each maximum dense subset $U_w$, the star edges are added to $E'$. The star edges and the edges of the implied dense subset graph $E(U_w)=\{\{u,v\}\in E | u,v\in U_w\}$, which are the covered edges, are removed from $E$, so that the covered edges are simply entirely removed. Since we have to calculate the density of the maximum density subset, we have $E(U_w)$ already available, so no extra work is required to calculate the edges induced by the subset of vertices.

\mypar{Berman et al.\ (\BBMRY).}
The original algorithmic description leaves some implementation details open. First, the property of an arc $(s,t)$ being \emph{thin} or \emph{thick} has to be efficiently evaluated \cite[Definition 2.2]{Berman2013}. The definition uses a \emph{local graph} $G^{s,t}=(V^{s,t},E^{s,t})$ induced by all vertices belonging to paths from $s$ to $t$ with a length of at most $\alpha\cdot d_G(s,t)$. An arc~$(s,t)$ is thick, if $|V^{s,t}|\geq n^{1/2}$. Calculating all possible paths is not an option, so in- and out-arborescenses can be used.
First, we precalculate these arborescences since we need them for evaluating the local graphs and during sampling afterwards. For each vertex $r$, an in- and out-arborescense rooted at $r$ is calculated. The distances to other vertices $x$ are saved in $d_\mathrm{in}(r, x)$ ($d_\mathrm{out}(r, x)$, respectively) together with the respective predecessor to be able to access the actual trees during sampling. To check if $|V^{s,t}|\geq n^{1/2}$ holds, we can use the equivalence: $x\in V^{s,t} \Longleftrightarrow d_\mathrm{out}(s, x)+d_\mathrm{in}(t, x)\leq\alpha\cdot d_G(s,t)$. Checking each vertex $x$, we count $|V^{s,t}|$ and decide the property for each arc. This could be done by using either an in- or out-arborescence, but both types are needed for the sampling, so we can precalculate both anyhow. Consequently, the subsequent sampling does not involve any shortest-path calculations and can be done by looking up predecessor relations.

Next, given an arc $(s,t)$ it must be checked, if an set $R\subseteq E$ \emph{settles} the arc \cite[Definition 2.3]{Berman2013}, i.e., $R$ satisfies the $\alpha$-spanner property for $(s, t)$. This is done by running a bounded Dijkstra on $R$ with a running time of $\O(|R|+n\log n)$. The check whether an arc is settled is done quite often, e.g., during the separation method or minimizing of an antispanner.

Finally, let us focus on the separation method and the creation of antispanners. The former is straightforward: After rounding the fractional solution, obtaining $R$, we check every thin arc, whether it is settled by $R$. For an unsettled thin arc $(s,t)$, we create an antispanner w.r.t.\ $R$ (see \cite[Claim 2.4]{Berman2013}). Similar to the thin arc property where the vertices $V^{s,t}$ are required, we need the arc set $E^{s,t}$, as $A=E^{s,t}\setminus R$ is an antispanner. An arc $(u,v)$ is an element of $E^{s,t}$ if and only if $d_\mathrm{out}(s, u)+w(u,v)+d_\mathrm{in}(t, v)\leq\alpha\cdot d_G(s,t)$; this check can reuse the precalculated arborescences. After finding the antispanner it has to be minimized. We greedily remove arcs from $A$, while the unsettled thin arc remains unsettled. If no further arc can be removed, $A$ is a minimal antispanner.

\mypar{Baswana and Sen (\BS).}
We broadly follow the detailed algorithmic description of \cite{Baswana2007}. We changed only one small aspect: The second and third step of the first phase can be combined in a single pass. After sampling clusters, it is not necessary to first calculate the nearest neighboring sampled cluster for each vertex and store it in some data structure because the third step can directly follow and does not depend on other vertices.

\mypar{Elkin and Neiman (\EN).}
Provided the very compact description of the algorithm and no explicit pseudocode, one has to be very careful not to gloss over intricate important details. We implemented the algorithm in the standard centralized model. The authors also highlight the distributed and PRAM models because the algorithm can be parallelized well, but we aim to keep it comparable to all other algorithms.

To start, remember that the edges saved together with the received messages must belong to \emph{some} shortest $u$-$x$-path. Naturally, we store the edge where $x$ received the message from. To ensure that the messages travel along shortest paths, we use a breath-first-search (BFS) starting at $u$ with a depth limit of $k$, without the need of a Dijkstra computation.

The algorithm includes two feasibility checks. If one fails, the algorithm does not provide a feasible solution. The first check, that $r_u<k$ holds for all $u\in V$, can be done directly while generating values $r_u$ to let the algorithm fail quickly if the property does not hold. The second check at the end of the algorithm asserts that there are sufficiently enough edges. Elkin and Neiman state that the spanner must have at least $n-1$ edges and the algorithm fails, if $|E'|<n-1$. This is only correct if the original graph is connected. If it has $c\geq2$ components, we can use the condition $|E'|\geq n-c$ instead.

A major improvement in terms of memory consumption can be made when reversing the way the algorithm is formulated. The original formulation can chiefly be described as:
\begin{enumerate}
	\item For each $u\in V$, broadcast messages $m_u(x)$ to every vertex $x$ within distance $k$.
	\item For each receiving $x\in V$: Calculate $\max_{u\in V}\{m_u(x)\}$ and add edges to the spanner.
\end{enumerate}
Following this order, one has to save two $n\times n$ matrices containing the message values $m_u(x)$ and the edge where the messages are received. In a pilot study, this resulted in out-of-memory errors for very sparse graphs ($m\approx2n$) with $n\approx 40,000$, which other algorithms can handle. To lower the memory consumption, we reversed the logic to focus on each receiving vertex $x$: A BFS starting at $x$ identifies all vertices $u$ that can broadcast a message to $x$. Then, the aggregation and edge addition can directly be done afterwards, before proceeding with the next $x$. To summarize all improvements, the code is provided in the appendix.

We note that it is possible to provide values for $\varepsilon$ that are not in the interval $(0,1)$. Especially $\varepsilon\geq1$ can produce good spanners in practice with the downside of an increasing possibility to fail. For results of a pilot study to choose $\varepsilon$, see \cref{subsec:random}.

\section{Experimental Results}\label{sec:experimental-results}
In this section the main results are presented. The considered algorithms were tested with a variety of instances and parameter settings, see below. In \cref{subsec:random}, we focus on \BS and \EN to derive an understanding of their randomized behavior and their multi-run variants. Additionally, we need to find a good value for $\varepsilon$ to compare \EN to other algorithms. In \cref{subsec:runtime} we take a look at the running time of each algorithm and set a timelimit for the subsequent experiments. Finally, we compare the quality of the resulting spanners in \cref{subsec:quality}.

\mypar{Setup.}
Since \BBMRY is the only algorithm that would work on directed graphs, we consider only undirected graphs. Edge weights are often a requirement in applications, but not all algorithms can take weighted graphs as inputs; thus we use each graph twice, once with and once without weights. In the following we will use the absolute density $\varrho_a(G)=m/n$ and the relative density $\varrho_r(G)=m/{n\choose 2}$ to categorize graphs. In the experiments we consider integer stretches 2, 3, 4, 5, and 7. To test a variety of graph types, three instance libraries are used:
\begin{enumerate}
	\item \emph{Random}: We generated random Erd\H{o}s-R\'enyi graphs with $|V|\in\{10,\allowbreak 20,\allowbreak 50,\allowbreak 100,\allowbreak 200,\allowbreak 500,\allowbreak 1000,\allowbreak 2000\}$, a relative density of $i/10$ for $i\in\{1,2,\ldots,9\}$ and ten graphs of each combination. The graphs are edge weighted with random integer weights between $1$ and $n$. These 720 graphs are available at \url{http://tcs.uos.de/research/spanner}.
	\item \emph{Steinlib} \cite{Steinlib}: This well-known library of graphs for the Steiner tree problem contains 1207 graphs in 44 subsets. Over 62\% of the graphs have $\varrho_r<2.5$\% (mostly sparse).
	\item \emph{Tsplib} \cite{Tsplib}: All 122 instances of the Tsplib are complete weighted graphs. We omitted 27 graphs that have edge weights 0 or are directed, resulting in 95 considered graphs. About half of the graphs have less than 200 vertices and the largest graph has 7397 vertices.
\end{enumerate}

All experiments were performed on an Intel Xeon Gold 6134 with 256 GB RAM. All algorithms are implemented as part of the open source C++ \emph{Open Graph Algorithms and Datastructures Framework} \cite{OGDF} (\url{www.ogdf.net}) and are compiled with the \texttt{-O3} flag. For \BBMRY we use CPLEX 20.1 \cite{CPLEX} as the LP solver. All detailed data of our experiments are available at \url{http://tcs.uos.de/research/spanner}.

\subsection{Parametrizing the Randomized Algorithms}\label{subsec:random}
To fairly evaluate the randomized algorithms \BS and \EN, we first need to investigate how often they should be run per instance until a \emph{good} result can be expected. To estimate the number of required iterations, we run each algorithm many times on instances to get a large sample size per instance. We consider two measures defining how many of the samples are \emph{good}. The spanner’s size is used to measure the quality with a lower value being better. For a given instance, let $\mathcal{X}$ be the samples and $\bar{\mathcal{X}}$ the mean of the samples' spanner sizes:
\begin{description}
	\item[M1] The number of runs that yield a size less than $1.25\cdot\min \mathcal{X}$.
	\item[M2] The number of runs that yield a size less than $0.25\cdot(\bar{\mathcal{X}}-\min{\mathcal{X}})+\min{\mathcal{X}}$.
\end{description}
The first measure is motivated by $\alpha$-approximations and is sensitive to the absolute values of the samples. This motivates the second measure, which is independent of the absolute sizes of the spanners. It counts the number of values in the lower 25\% between the lowest and mean value. We performed 100 batches with $10{,}000$ runs for \BS (1000 runs for \EN due to higher running time) for each instance configuration. Due to the computational effort, we consider six random instances with increasing relative density. Each instance is evaluated with and without edge weights and with $\alpha=3$ and $\alpha=7$.

Before evaluating \EN, we have to find a value for $\varepsilon$ that we will use in the following.
% random/50/30.gml
We tested values for $\varepsilon$ from $0.01$ up to $3$. For all instances and stretches the behavior was identical: The number of failed iterations increases with increasing $\varepsilon$, but more strongly for $\varepsilon\geq2$. The mean value of all valid samples decreases with higher values for $\varepsilon$ but stagnates around $\varepsilon\geq0.8$. Additionally, the standard deviation also decreases with higher $\varepsilon$. We thus fixed $\varepsilon=0.8$ as it offers the best compromise between objective value, deviation and failure rate. The high deviation allows for sampling lower values with a higher probability. The failure rate for $\varepsilon=0.8$ is just below 25\%. In the experiments for \cref{subsec:runtime}, the rate of success on first tries is slightly above 75\%, in each of the three libraries.

\begin{table}[t]
	\setlength\extrarowheight{1pt}
	\centering
	\smaller
	\begin{tabular}{|l|c|c|c|c|}\hline
		 & M1, sparse & M1, dense & M2, sparse & M2, dense \\ \hline\hline
		\BS, unweighted & 10 & 500 & 700 & 100 \\
		\BS, weighted & 10 & 10 & 700 & 100 \\
		\EN, unweighted & 5 & 200 & 150 & 20 \\ \hline
	\end{tabular}
	
	\caption{Calculated number of required iterations for M1 and M2.}
	\label{table:required-iterations}
\end{table}

\cref{table:required-iterations} provides an overview for each measure w.r.t.\ the number of required iterations. The behavior with respect to the relative density is linear, so we only give the extreme values for sparse graphs ($\varrho_r<10\%$) and dense graphs ($\varrho_r>90\%$). A Shapiro-Wilk test shows that the samples are not normally distributed. The best number of required iterations in \cref{table:required-iterations} does not depend on the stretch, but behave inversely for M1 and M2. For unweighted sparse graphs, the resulting spanner sizes have a low variance in contrast to dense graphs, so M1 counts more good iterations for sparse graphs. The spanner sizes are generally larger for weighted graphs, so that M1 counts many samples most of the time, resulting in an overall low number of required iterations. Interestingly, the excess kurtosis and the skewness of the samples increases with a higher density (cf.~\cref{fig:bs-random} in the appendix). Both increasing moments indicate more compact data with the balance of the distribution shifted towards the lower values, so M2 contains more good iterations for high density graphs since more iterations are captured in the lower 25\% interval.

Overall, multiple iterations are worthwhile, but the "best" reasonable number varies depending on the structure of the graph  and the application. For the experiments in \cref{subsec:quality}, we limit \BS to 1000 iterations and \EN to 200 iterations, since generally less iterations are required by \EN.

\subsection{Running time}\label{subsec:runtime}
We now take a look at the running time performance of our algorithms. We set a timelimit of 60 seconds for each calculation since \ADDJS, \BS, and \EN can solve a majority of instances within this limit (cf.~\cref{table:success}). Additionally a pilot study shows, that higher limits (90, 120 seconds) does not significantly increase the number of solved instances for all algorithms. First, we discuss them individually, before comparing them afterwards.

% runtime_stats.py
\begin{table}[t]
	\setlength\extrarowheight{1pt}
	\centering
	\smaller
	\begin{tabular}{|@{\ }c@{\ }|@{\ }c@{\ }|@{\ }c@{\ }|@{\ }c@{\ }|@{\ }c@{\ }|@{\ }c@{\ }|}\hline
		Library & \ADDJS & \KP & \BBMRY & \BS & \EN \\ \hline\hline
		Random & $99.05\%$ ($2.96$~s) & $66.67\%$ ($2.46$~s) & $71.11\%$ ($2.53$~s) & $100.00\%$ ($0.03$~s) & $100.00\%$ ($3.10$~s) \\
		Steinlib & $99.88\%$ ($0.48$~s) & $91.55\%$ ($1.91$~s) & $83.79\%$ ($5.32$~s) & $100.00\%$ ($0.11$~s) & $100.00\%$ ($0.37$~s) \\
		Tsplib & $81.75\%$ ($4.98$~s) & $33.01\%$ ($14.75$~s) & $60.74\%$ ($5.70$~s) & $98.71\%$ ($1.47$~s) & $76.05\%$ ($3.11$~s) \\ \hline \hline
		All & $98.48\%$ ($0.87$~s) & $86.02\%$ ($2.30$~s) & $81.19\%$ ($5.23$~s) & $99.90\%$ ($0.21$~s) & $98.20\%$ ($0.66$~s) \\ \hline
	\end{tabular}
	\caption{Solved instances per algorithm and average time for successful cases.}
	\label{table:success}
\end{table}

\mypar{\ADDJS.} The running time is linear with $m$ for a fixed $n$. Almost all instances of the Steinlib can be solved within 6 seconds and all instances from the Tsplib with $n\leq1200$ within the timelimit. The densest instances over all libraries with $\varrho_r\geq90\%$ and $n\approx1000$ can barely be solved within the timelimit. The running time shows a dependency to graph weights: The unweighted graph versions of Random can be solved $33.67$\% faster than the weighted ones. Additionally, a higher stretch results in a faster running time. Setting $\alpha=2$ for unweighted graphs has an outstandingly high running time to all other combinations of weights and stretch: For Random, only 90\% of instances can be solved while all other combinations yield a 100\% solve rate. For those instances which can be solved, the running time is $42.61$\% higher as $\alpha=3$ (in comparison: the running time for $\alpha=3$ is only $6.19$\% higher than $\alpha=4$ and for $\alpha=4$ it is only $13.03\%$ higher than $\alpha=5$).

\begin{figure}[t]
	\begin{minipage}{.48\textwidth}
		\centering
		\hspace*{-0.4cm}
		\includegraphics[width=1.1\linewidth]{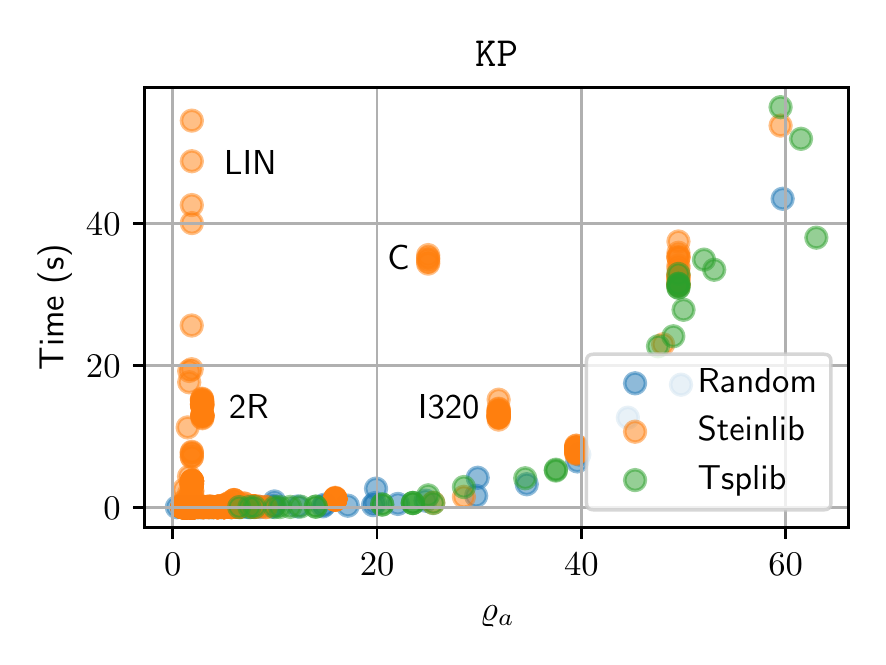}
		\vspace*{-1cm}
		\captionof{figure}{Instances solved by \KP for each instance library. Instances with exceeded timelimits are excluded.}
		\label{fig:kp-runtime}
	\end{minipage}
	\hfill
	\begin{minipage}{.48\textwidth}
		\centering
		\hspace*{-0.5cm}
		\includegraphics[width=1.1\linewidth]{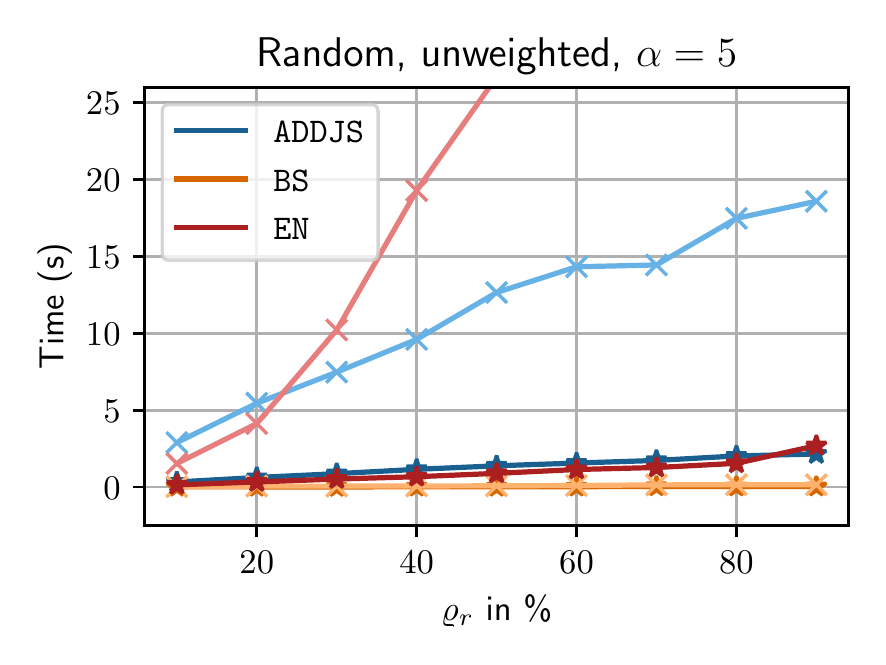}
		\vspace*{-1cm}
		\captionof{figure}{Comparing \ADDJS, \BS, and \EN for $n=500$ ($\times$) and $n=1000$ (*). \BBMRY cannot solve any of these instances.}
		\label{fig:runtime-comparison}
	\end{minipage}
\end{figure}

\mypar{\KP.} Surprisingly, there is no clear relationship between $n$, $m$, or $\varrho_r$ to the running time; however the absolute density is limiting: For instances with $40\lesssim\varrho_a\lesssim60$, the running time drastically starts to increase in comparison to $\varrho_a\lesssim40$ and no instances with $\varrho_a\gtrsim60$ can be solved in time (cf.~\cref{fig:kp-runtime}). This behavior is consistent over all three libraries except for a handful of outliers from Steinlib. Regarding Steinlib, the sets \texttt{2R}, \texttt{I320}, \texttt{C}, \texttt{PUC} and the largest instances of \texttt{LIN} have higher running times due to the very large $n$. The complete graphs of Tsplib can be solved with up to 120 vertices, which corresponds to $\varrho_a=59.5$.

\mypar{\BBMRY.} Its running time depends on $\alpha$ and graph weights: Similar to \ADDJS, creating spanners for unweighted graphs is slightly faster by $9.52$\%, but a higher stretch results in a lower running time. Also similar is the varying behavior for $\alpha=2$ and unweighted graphs. Over all, both the timeout rate and the running time in the successful cases are very high. Regarding the instance sizes, the number of edges has a disproportionate impact on the running time, i.e., only Steinlib instances with $n\leq5000$ edges can be solved, regardless of the number of vertices. This is not surprising since the antispanner creation and especially the minimization of each antispanner is an operation that depends on the original graph's edges. Graphs from Random and Tsplib can only be solved if they have less than 250 nodes. Even though \BBMRY has a probability of failing, we never had any failure during the experiments.

\mypar{\BS.} The clustering approach is very fast. Every instance from Random is solved in under $0.6$ seconds. The running time is linear in $m$ as expected. Analogous to the algorithms above, unweighted instances are faster, but in contrast, a lower stretch is faster as well. Almost all instances of the Tsplib can be solved; only the two largest instances ($n=5915$ and $n=7397$) exceed the timelimit for $\alpha\geq5$. The majority of instances from the Steinlib can be solved within one second. Interestingly, \BS seems to have difficulties for very sparse and large graphs; For these graphs with $\varrho_r\leq0.01\%$, the running time is up to $17.5$ seconds (cf.~\cref{fig:bs-runtime} in the appendix). These graphs are the larger instances from the \texttt{LIN}, \texttt{ALUT}, and \texttt{ALUE} sets of Steinlib, which are large gridgraphs ($n\geq20{,}000$) with holes. 

\mypar{\EN.} It shows a running time behavior linear in $m$. We measured the time until the first success, so previous failures are included in the running times. It can solve graphs with up to 1000 vertices for Random and Tsplib, with a smaller stretch being slightly faster. The maximum time for Steinlib is 14 seconds for instances with many vertices.

\mypar{Comparison.}
After describing the algorithms itself, we are going to compare them. \cref{table:runtime-instances} shows the percentage of instance one algorithm can solve strictly faster than another algorithm. Additionally, the average factor is given, how much faster the algorithm is. An example is provided in \cref{fig:runtime-comparison} for unweighted instances and $\alpha=5$. It is clear that, except for \KP in only 8.2\% cases, all algorithms are faster than \BBMRY and \BBMRY is only 14 times faster than \KP. However, if \KP is faster, it is on average 116 times faster than \BBMRY. \BS shows a significant number of instances where it is faster, because it benefits from traversing adjacency lists in contrast to calculating distances. Also the running time factor to other algorithms is comparatively high. \EN and \ADDJS both share a similar amount of instances each can solve faster. This is no surprise since the BFS for every node is similar to the Dijkstra runs for \ADDJS in the unweighted setting. Only for large or dense instances \ADDJS gets noticeably faster than \EN (cf.~\cref{fig:runtime-comparison}). If these values would be restricted to larger or denser graphs (e.g. $n\geq100$ or $\varrho_r\geq5\%$), the results would be even clearer due to the fact that all small/sparse instances can be solved within fractions of seconds by most algorithms.

\begin{table}[t]
	\setlength\extrarowheight{1pt}
	\centering
	\smaller
	\begin{tabular}{|c|c|c|c|c|c|}\hline
		\backslashbox{A}{B} & \ADDJS & \KP & \BBMRY & \BS & \EN \\ \hline
		\ADDJS & \cellcolor{black!10} & $51.23\%$ ($123.61$) & $93.78\%$ ($212.16$) & $15.18\%$ ($1.45$) & $19.38\%$ ($1.07$) \\ \hline
		\KP & $1.19\%$ ($1.27$) & \cellcolor{black!10} & $86.65\%$ ($116.61$) & -- & -- \\ \hline
		\BBMRY & $0.00\%$ (--) & $8.20\%$ ($14.17$) & \cellcolor{black!10} & $0.00\%$ (--) & $0.00\%$ (--) \\ \hline
		\BS & $35.57\%$ ($23.58$) & -- & $93.53\%$ ($189.70$) & \cellcolor{black!10} & $28.70\%$ ($30.23$) \\ \hline
		\EN & $27.10\%$ ($2.68$) & -- & $93.37\%$ ($170.12$) & $11.52\%$ ($0.45$) & \cellcolor{black!10} \\ \hline
	\end{tabular}
	\caption{Percentage of instance where A was strictly faster than B. "--" denotes that the algorithms cannot consider common instances due to their stretch restrictions (2 vs. odd), see \cref{table:algorithms}. The speed up factor is given in parenthesis. Only instances both algorithms can solve are included.}
	\label{table:runtime-instances}
\end{table}

\subsection{Quality}\label{subsec:quality}
Finally, we take a look at the resulting lightness and sparseness of the spanners. We also take a look at the spanners' node degrees and the effective stretch realized by the algorithms. We also take a quick look at the number of edges of each shortest path in the weighted case.

\begin{figure}[t]
	\begin{minipage}{.48\textwidth}
		\centering
		\hspace*{-0.4cm}
		\includegraphics[width=1.1\linewidth]{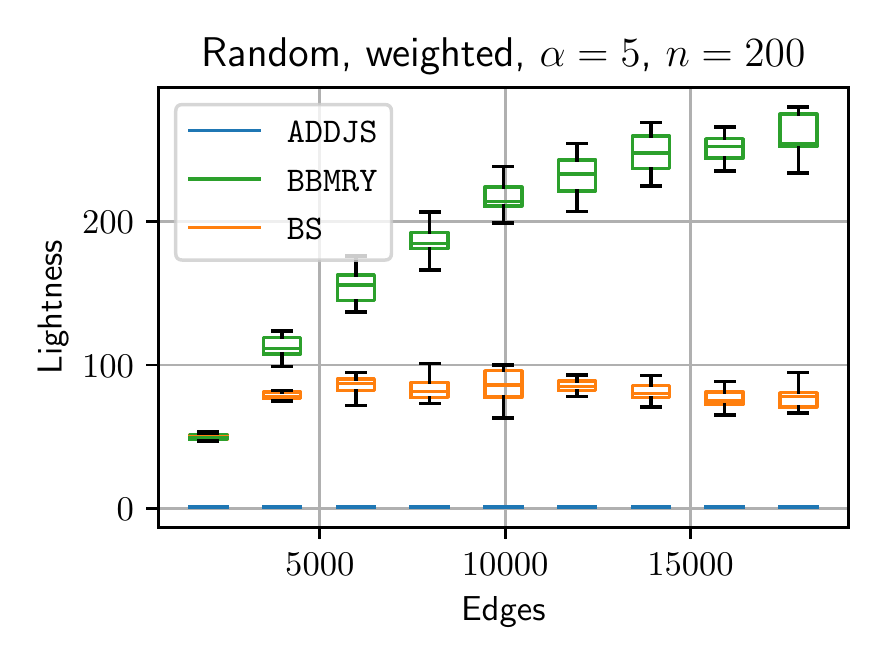}
		\vspace*{-1cm}
		\captionof{figure}{Lightness for \ADDJS, \BBMRY and \BS. \ADDJS has a maximum lightness of $1.3$.}
		\label{fig:lightness}
	\end{minipage}
	\hfill
	\begin{minipage}{.48\textwidth}
		\centering
		\hspace*{-0.5cm}
		\includegraphics[width=1.1\linewidth]{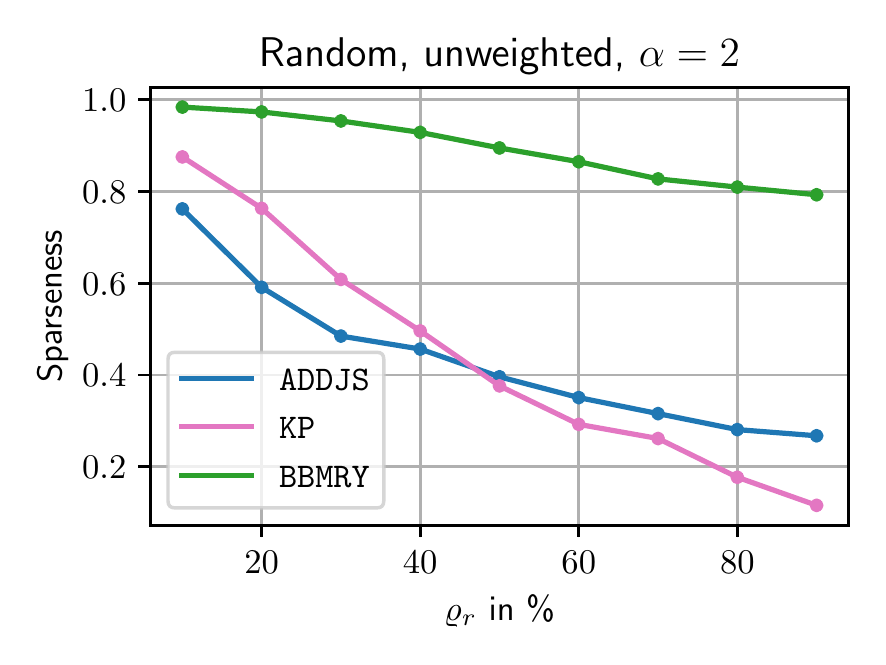}
		\vspace*{-1cm}
		\captionof{figure}{Sparseness comparison averaged on all intances from Random.}
		\label{fig:sparseness-kp}
	\end{minipage}
\end{figure}

\mypar{Lightness.} As a concept different from size, the lightness is only relevant for weighted instances. \BBMRY and \BS do not provide any guarantees, and indeed, both yield arbitrarily high lightness values in practice across all instance libraries (cf.~\cref{fig:lightness}). \BBMRY's lightness is on average $33.5$ times higher than \ADDJS and \BS's lightness is $11.8$ times higher. \BS and \BBMRY never yield a lower lightness than \ADDJS on any instance. In summary, there is no alternative to \ADDJS for applications requiring a lightness guarantee.

\mypar{Sparseness.} We made a surprising observation for $\alpha=2$ and unweighted graphs. For relative densities above roughly 50\%, \KP yields a lower sparseness than \ADDJS (cf.~\cref{fig:sparseness-kp}). \BBMRY yields only high sparseness in comparison to those algorithms. On Tsplib (i.e., complete graphs), \KP always finds the trivial optimal solution of a star due to the way the algorithm works. In contrast, \KP can only sparsify 22\% of the Steinlib instances in comparison to 78\% by \ADDJS and 62\% by \BBMRY. Similar to the high running time for \ADDJS with unweighted graphs and $\alpha=2$, the sparseness is 137\% higher than $\alpha=3$, while $\alpha=3$ has only a 26\% higher sparseness than $\alpha=4$. Since we do not know the optimal solution values, this may still be a good value, but for dense graphs, \KP results in lower sparseness.

\begin{figure}[t]
	\begin{minipage}{.48\textwidth}
		\centering
		\hspace*{-0.4cm}
		\includegraphics[width=1.1\linewidth]{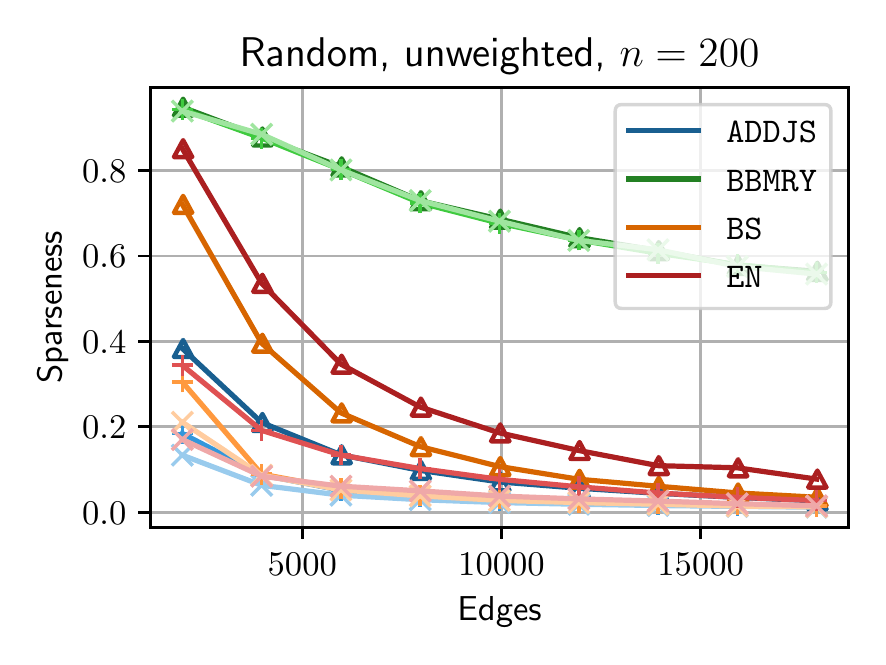}
		\vspace*{-1cm}
		\captionof{figure}{Sparseness for unweigted graphs with different stretches: $\alpha=3$ ($\triangle$), $\alpha=5$ ($+$), and $\alpha=7$ ($\times$).}
		\label{fig:sparseness-unweighted}
	\end{minipage}
	\hfill
	\begin{minipage}{.48\textwidth}
		\centering
		\hspace*{-0.5cm}
		\includegraphics[width=1.1\linewidth]{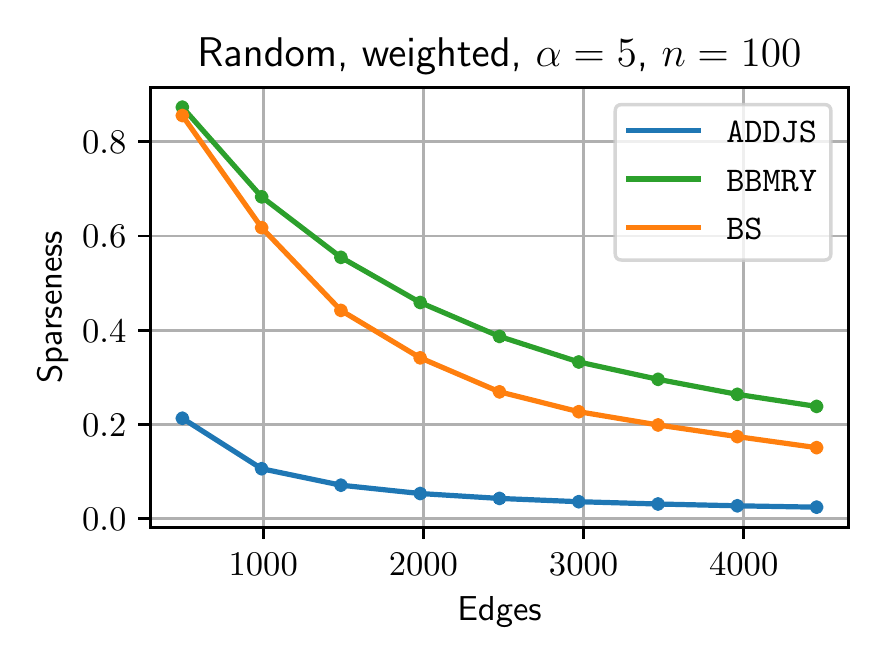}
		\vspace*{-1cm}
		\captionof{figure}{Sparseness for weigted graphs. \BBMRY and \BS are significantly higher than \ADDJS.}
		\label{fig:sparseness-weighted}
	\end{minipage}
\end{figure}

For unweighted instances with odd stretch, \ADDJS, \BS, and \EN show similar behaviors to each others, see \cref{fig:sparseness-unweighted}. For dense graphs, the sparseness converges and higher stretches unsurprisingly result in lower sparsenesses. \ADDJS generally yields the lowest sparseness with \BS yielding a 37\% higher sparseness, and \EN having a 47\% higher sparseness than \BS (unweighted graphs, $\alpha=5$, all libraries). \BBMRY is special as its sparseness is stretch-independent; but it yields only an overall high sparseness compared to all other algorithms. \EN shows a higher variance of the sparseness for $\alpha=3$ with a standard derivation about 4 times larger in comparison to \ADDJS. With higher stretches, the variance decreases. On weighted graphs, \BBMRY and \BS offer similar sparsenesses, which are significantly higher than \ADDJS, see \cref{fig:sparseness-weighted}. All algorithms except for \ADDJS cannot significantly sparsify the already sparse graphs ($\varrho_r\leq10\%$) from Steinlib too much: \ADDJS's average sparseness is 71.7\%, while the sparseness of all other algorithms are between 92\% and 97\%.

\mypar{Degree.}
Comparing the mean degree of the spanner to that of the original graph, \ADDJS and \BS interestingly yields a fairly constant spanner degrees for unweighted graphs from Random. The mean spanner degree is in the range of $20\pm10$, depending on the instance. On average, \EN yields higher degrees as \ADDJS and \BS and as for the sparseness, it shows a high variance for $\alpha=3$ (cf.~\cref{fig:degree-random} in the appendix). \BBMRY halves the mean degree. The stretch does not have an influence for all algorithms. For $\alpha=3$, \BS yields a higher mean degree for weighted graphs by 30\% than \BBMRY, but is slightly lower for $\alpha\geq5$. \BS is the only algorithm where a higher stretch results in lower mean degrees for weighted graphs. \BBMRY and \BS yield very high mean degrees (similar to the lightness and sparseness) while \ADDJS has mean degrees around 3 to 5. Also similar to the lightness is the behavior for Steinlib instances: On average, \ADDJS has 57\% of the original mean degree while all other algorithms' spanners have more than 96\% of the original mean degree, see \cref{fig:degree-steinlib} in the appendix. All algorithms show a constant or moderately (linearly) increasing dependency between the spanner and original graph, except for \KP. First, the spanners' mean degrees increase together with the original mean degree, but for instances with $\varrho_r\gtrsim50\%$ they decrease again. Analogous to the sparseness, the mean degrees become lower than \ADDJS's for large relative densities.

\mypar{Effective Stretch.}
For instances with $n\leq1000$ ($n\leq11{,}000$ for Steinlib), we computed an APSP on the original graphs and the resulting spanners to calculate the effective stretches. \cref{table:quality-stretch} provides an overview of the mean stretch, the averaged max stretches and the highest stretch on all instances. \ADDJS gives the overall highest mean stretch and always utilizes the limit provided by $\alpha$. Interestingly, \BBMRY, \BS, and \EN never use the available stretch of 7 for unweighted graphs. For unweighted complete graphs, \EN does not yield a mean stretch over 2 (cf.~\cref{fig:stretches-tsplib} in the appendix), regardless of the allowed stretch. \KP behaves similar to \ADDJS and yielding even an higher mean stretch if \ADDJS is restricted to instances \KP could solve. \BBMRY and \BS yield lower mean stretches in comparison to \ADDJS for weighted graphs. In summary, these results directly correlate to the previous observations. By having a low mean stretch and not utilizing the allowed stretch, it is clear that some potential for sparsification is not fully exploited, so the lightness and sparseness are also worse.

\begin{table}[t]
	\setlength\extrarowheight{1pt}
	\centering
	\smaller
	\begin{tabular}{|c|c|c|c|c|c|c|}\hline
		& $\alpha$ & \ADDJS & \BBMRY & \BS & \KP/\EN \\ \hline
		\multirow{5}{*}{\rotatebox[origin=c]{90}{unweighted}} & 2 & $1.36$ ($1.98$, $2.00$) & $1.10$ ($1.86$, $2.00$) & -- & \KP: $1.36$ ($1.87$, $2.00$) \\
		& 3 & $1.77$ ($2.94$, $3.00$) & $1.12$ ($2.06$, $3.00$) & $1.53$ ($2.77$, $3.00$) & \EN: $1.43$ ($2.61$, $3.00$) \\
		& 4 & $2.16$ ($3.89$, $4.00$) & $1.13$ ($2.13$, $4.00$) & -- & -- \\
		& 5 & $2.47$ ($4.77$, $5.00$) & $1.13$ ($2.11$, $4.00$) & $1.77$ ($3.68$, $5.00$) & \EN: $1.59$ ($3.13$, $5.00$) \\
		& 7 & $2.86$ ($6.43$, $7.00$) & $1.13$ ($2.12$, $5.00$) & $1.83$ ($4.16$, $6.00$) & \EN: $1.73$ ($3.52$, $6.00$) \\ \hline
		\multirow{5}{*}{\rotatebox[origin=c]{90}{weighted}} & 2 & $1.03$ ($1.85$, $2.00$) & $1.00$ ($1.19$, $2.00$) & -- & -- \\
		& 3 & $1.07$ ($2.70$, $3.00$) & $1.00$ ($1.65$, $3.00$) & $1.01$ ($1.16$, $2.61$) & -- \\
		& 4 & $1.11$ ($3.49$, $4.00$) & $1.01$ ($1.99$, $4.00$) & -- & -- \\
		& 5 & $1.16$ ($4.20$, $5.00$) & $1.01$ ($2.30$, $5.00$) & $1.01$ ($1.37$, $3.26$) & -- \\
		& 7 & $1.25$ ($5.44$, $7.00$) & $1.01$ ($2.69$, $7.00$) & $1.01$ ($1.60$, $3.49$) & -- \\ \hline
	\end{tabular}
	\caption{Effective mean (mean max, max) stretch for multiple configurations.}
	\label{table:quality-stretch}
\end{table}

\mypar{Hops.}
Lastly, we take a look at the number of edges (hops) of all shortest paths for the same restricted instances as considered for the effective stretch. Hops are only interesting for weighted graphs since otherwise it is the same as the effective stretch. All results considered, the mean hop difference is always positive, so on average shortest paths in the spanner use more edges than in the original graph. Only $2.27\%$ of all vertex pairs have shortest paths with fewer hops in the spanner than in the original graph; $58.90\%$ have the same number of hops, and $38.82\%$ gain at least one more hop. FOr weighted graphs, \ADDJS has a higher average mean hop difference of $3.98$ than \BS ($0.26$) and \BBMRY ($0.15$). The tendency for all algorithms is to yield a higher hop difference with higher stretch.

\section{Conclusions}\label{sec:conclusions}
We conducted the first experimental evaluation of polynomial spanner approximations. Depending on the needs and settings of an application, we can provide a rough guideline which algorithm to use. If one has the special case of $\alpha=2$ and no edge weights, \KP may be used for graphs with approximately $\varrho_r\gtrsim50$ as long as it can solve the instances in feasible time. In all other cases \ADDJS (the oldest and most simplistic approach!) should be the algorithm of choice, as long as the graphs have reasonable sizes. It provides the sparsest and lightest spanners within a reasonable calculation time. Especially for weighted graphs, no other algorithm can calculate spanners of comparable quality. Only for unweighted very large graphs, \BS is a good alternative to \ADDJS due to its low running time. Using \BBMRY results in the worst quality spanners and has a disproportionate high running time. Only if directed graphs are considered, \BBMRY is the only algorithm to solve them. Lastly, \EN can produce good spanners, but it is never strictly better in running time nor quality than \ADDJS or \BS.

We may close with two questions: Can \ADDJS be improved by utilizing a specific order of edges, if the graph is unweighted? Currently, no order is specified if there are no edge weights, so an arbitrary implementation-specific order is used. Maybe using a BFS or DFS order, or sequentially considering sets of independent edges, can improve the spanners size even more in practice. It also may be interesting to further investigate the algorithmic reasons for the significant running time degradation of \ADDJS and \BBMRY for the special case of $\alpha=2$ on unweighted graphs.

\clearpage
\bibliography{literatur}

\clearpage
\appendix
\section{Pseudocode for \EN}

\begin{algorithm}[h]
	\SetKwFunction{Fail}{Fail}
	\SetKwFunction{NumberOfComponents}{NumberOfComponents}\SetKwData{Null}{null}
	\SetKwProg{Forall}{for all}{ do}{end}
	\SetKwInOut{Input}{Input}\SetKwInOut{Output}{Output}
	\Input{$G=(V, E)$, $\alpha\geq3\in\mathbb{N}_\mathrm{odd}$, $\varepsilon>0$}
	\Output{An $\alpha$-spanner $H=(V,E')$}
	
	$E'$ := $\emptyset$, $k$ := $(\alpha+1)/2$, $\beta$ := $\ln(3n/\varepsilon)/k$\;
	\Forall{$u\in V$}{
		$r_u$ := $\mathrm{EXP}(\beta)$\;
		\If{$r_u\geq k$}{
			\Fail\;
		}
	}
	\Forall{$x\in V$}{
		Queue $Q$\tcp*{Stores triple (vertex, depth, traveled edge)}
		Unmark all vertices, reset $m_y$ and $e_y$ for all $y\in V$\;
		$Q$.enqueue($x$, $0$, \Null)\;
		mark $x$\;
		\While{$Q$ is not empty}{
			$(y,distance,e)$ := $Q$.dequeue()\;
			\Forall{unmarked neighbors $u$ of $y$} {
				\If{$e=\Null$}{$e$ := $\{y, u\}$\;}
				
				$m_u$ := $r_u-distance$\;
				$e_u$ := $e$\;
				\If{$distance<k$} {
					$Q$.enqueue($u$, $distance+1$, $e$)\;
					mark $u$\;
				}
			}
		}
		\Forall{$u\in V$ with $m_u\geq\max_{n\in V}\{m_n\}-1$} {
			$E'$ := $E'\cup\{e_u\}$\;
		}
	}
	\If {$E'<(n-\NumberOfComponents(G))$} {
		\Fail\;
	}
	
	\caption{\EN algorithm. The $\beta$-parameterized exponential distribution $\mathrm{EXP}(\beta)$ has a probability density function $f(x)=\beta\cdot e^{-\beta x}$ for $x\geq 0$ and $f(x)=0$ otherwise.}
\end{algorithm}

\clearpage
\section{Further diagrams for the experimental evaluation}

\begin{figure}[h]
	\centering
	\includegraphics[width=\linewidth]{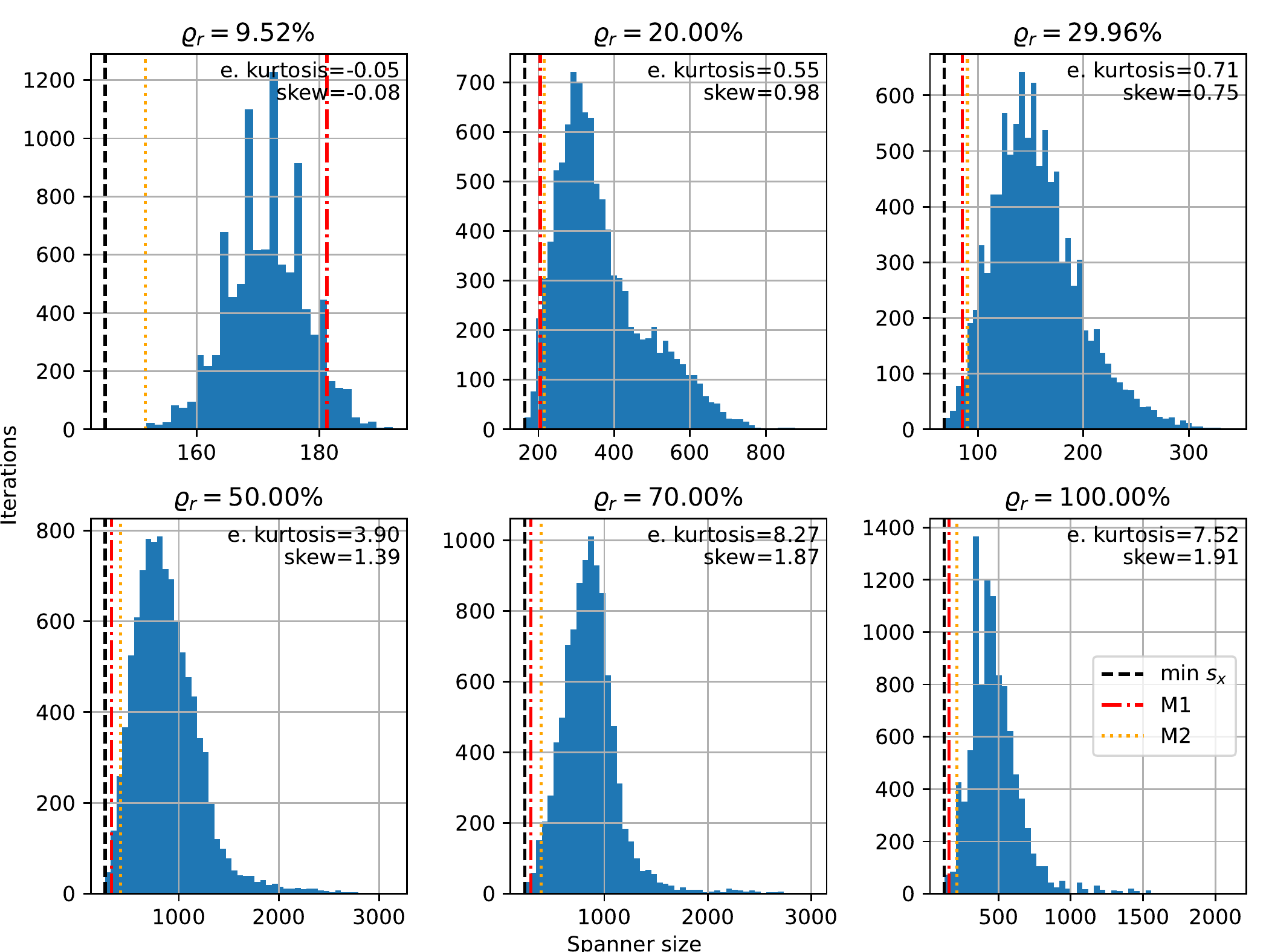}
	\caption{Six graphs with $10{,}000$ samples each, calculated with \BS for $\alpha=7$ and no edge weights. The black line indicates the minimum value, the red line the threshold for M1 and the orange line the threshold for M2.}
	\label{fig:bs-random}
\end{figure}

\begin{figure}[h]
	\begin{minipage}{.48\textwidth}
		\centering
		\hspace*{-0.4cm}
		\includegraphics[width=1.1\linewidth]{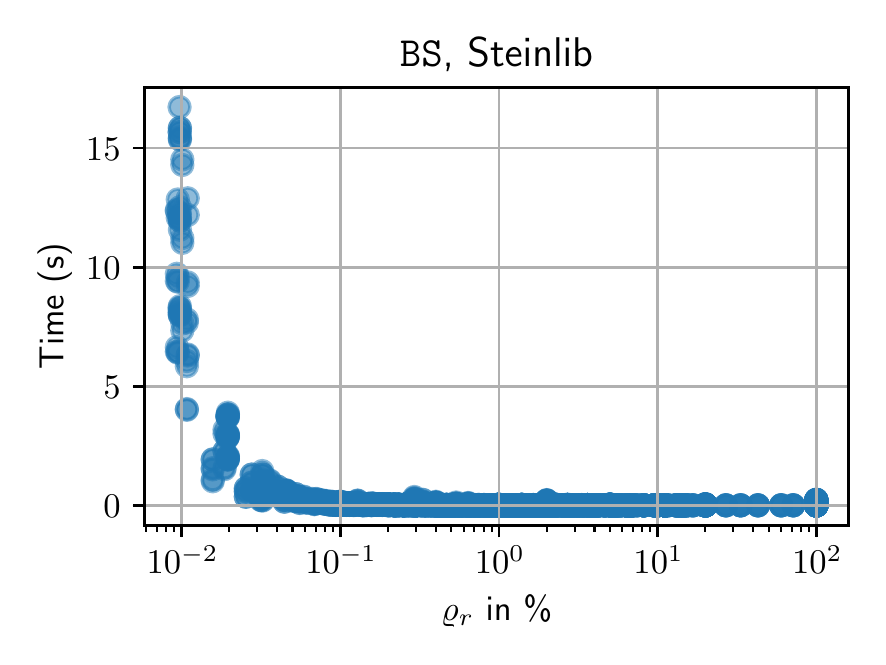}
		\vspace*{-1cm}
		\captionof{figure}{Results for \BS plotted by the relative density. A very low density results in a large running time.}
		\label{fig:bs-runtime}
	\end{minipage}
	\hfill
	\begin{minipage}{.48\textwidth}
		\centering
		\hspace*{-0.5cm}
		\includegraphics[width=1.1\linewidth]{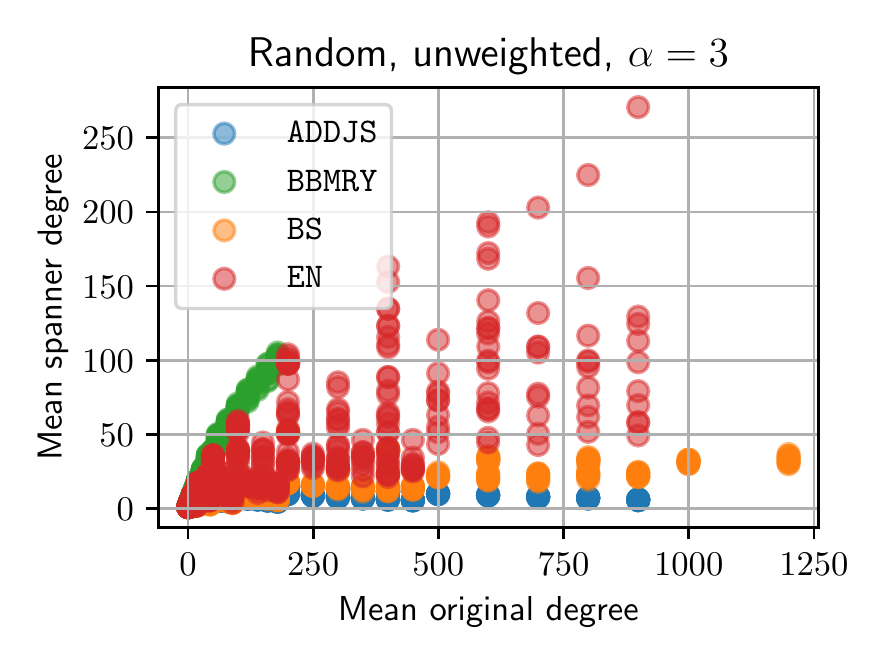}
		\vspace*{-1cm}
		\captionof{figure}{Spanner mean degree for the Random libraray. \EN shows a high scattering while all other algorithms have a linear dependency.}
		\label{fig:degree-random}
	\end{minipage}
\end{figure}

\begin{figure}[h]
	\begin{minipage}{.48\textwidth}
		\centering
		\hspace*{-0.4cm}
		\includegraphics[width=1.1\linewidth]{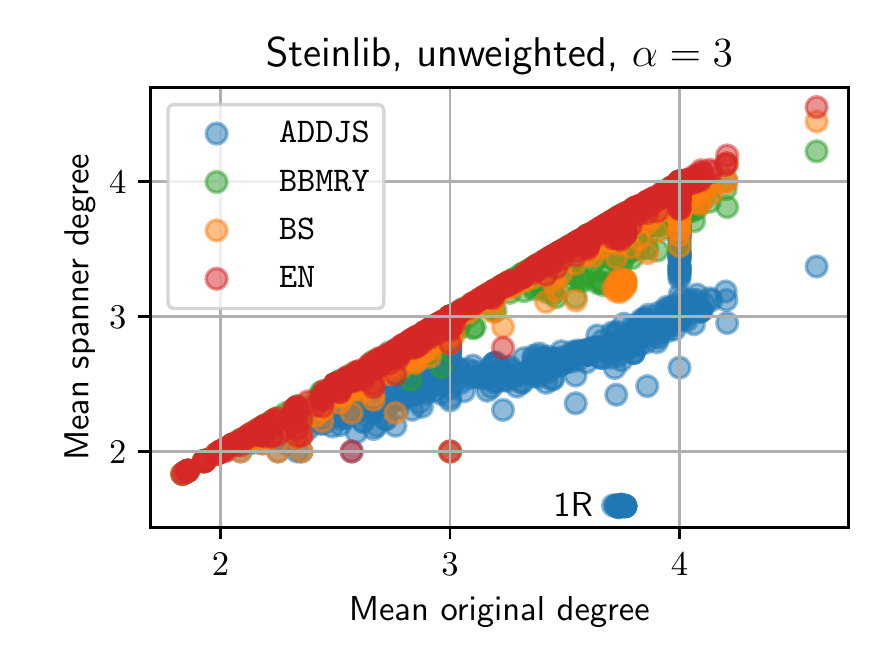}
		\vspace*{-1cm}
		\captionof{figure}{Spanner mean degree for the steinlib restriected to instances with an original degree less than 5.}
		\label{fig:degree-steinlib}
	\end{minipage}
	\hfill
	\begin{minipage}{.48\textwidth}
		\centering
		\hspace*{-0.5cm}
		\includegraphics[width=1.1\linewidth]{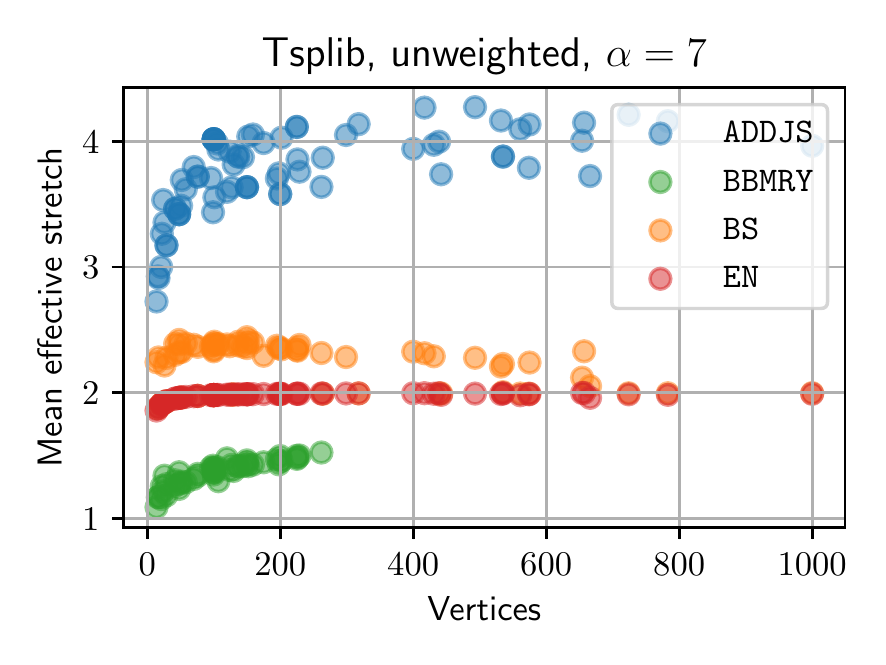}
		\vspace*{-1cm}
		\captionof{figure}{Mean effective stretches for unweighted complete graphs. \EN does not have a mean stretch over 2.}
		\label{fig:stretches-tsplib}
	\end{minipage}
\end{figure}

\end{document}